\documentstyle[prd,aps,epsfig]{revtex}

\begin{document}

\topmargin-1.0cm
\textheight21.5cm
\textwidth16cm
\oddsidemargin-0.5cm

\title{Nonrelativistic Quark Model Calculation of the $S$ parameter}

\author{Carlos A. Ramirez \footnote{Electronic address: cramire@uolpremium.net.co}}
\address{Escuela de F\'\i sica,\\
Universidad Industrial de Santander,\\
A. A. 678, Bucaramanga, Colombia}
\maketitle

\begin{abstract}
A critical review is presented of the attempts to estimate the Strong Interactions contributions to the parameter $S$ ($L_{10}$ in the QCD Chiral Version). In particular it is discussed why the estimations done for  Technicolor are unreliable.
$S$ is calculated for heavy doublets of Techniquarks using the Nonrelativistic Quark Model and keeping $v\simeq 0.25$ TeV fixed. It is found that heavy Techniquarks decouple, so it is possible to obtain values for $S$ in agreement with present experimental data.\\

\noindent PACS number(s): 12.60.Nz, 12.60.Fr, 12.38.Lg, 12.38.Aw, 12.39.Jh
\end{abstract}




\section{Introduction}

Parameters like $S$ have a twofold interest: first in the case of the Physics Beyond the Standard Model (SM) \cite{peskin,N_f,experiments,pdg} where it is used to account for deviations from the Standard Model predictions due to possible new Physics, and second in the case of QCD where $S_{\rm QCD}=-16\pi L_{10}$ and $L_{10}$ is the QCD Chiral Parameter \cite{gasser}.
Calculations for simple QCD models agree roughly with the experimental values \cite{cqm,vmd}, but when these results are scaled to obtain Technicolor (TC) \cite{tc,etc,wtc} theoretical predictions \cite{peskin,N_f,scaling,acd,dse,bse,gnc,ENJL}
conflict with experimental results arise and one may conclude that TC models are excluded (although people has been able to find several models where $S$ is negative \cite{negative-S}).
However many critics \cite{critics,isospin}  are unavoidable to the methods used to obtain $S$ in the cases where Strong Interactions are present, in particular when extrapolations are done to obtain TC predictions by scaling QCD results due to our poor understanding of nonperturbative QCD contributions. Let us review shortly these estimations and the critics they have received.
 
The first attempt to compute $S$ in the case of TC Models was done by scaling the QCD results \cite{peskin}. Although this estimation is reviewed in section II let us mention that one of the defects of this approach is the fact that the results are, once scaled independent of the Strong Interactions themselves! (the result is independent of the coupling constant and of the quark or Techniquark masses).
The same result is obtained by using the very simple Quark Model \cite{cqm} with constituent free quarks. In this model interactions are not considered and the result is essentially independent of the quark masses.

Another attempt was done by using Phenomenological Lagrangians \cite{gasser,N_f} in order to compute the pseudogoldstone mesons contributions, in the same way this is done in the case of the low energy QCD \cite{gasser}. Critics in this case arise because if the number of degenerate technidoubles, $N_{\rm TD}$ is large the masses of the resonances and the symmetry breaking scale, $\Lambda_\chi \sim 4\pi F_\pi/\sqrt{N_{\rm TD}}$ \cite{N_f} become too low and the energy expansion is not valid anymore \cite{N_f}.
Besides in Walking Technicolor (WTC) \cite{wtc} the Technipions may not be the Goldstone Bosons ($m_{\pi_{\rm TC}} \sim \Lambda_{\rm TC}$). Finally the result  $S\sim N_{\rm TD}^2$ disagrees with the other estimations where $S\sim N_{\rm TC}N_D$ because these two terms represent very different physics \cite{scaling}, the low and high energy contributions so they have to be added.
The Analytic Continuation by Duality (ACD) method \cite{acd} was also used but its confidence level is not very clear \cite{acd}. In this method one computes mainly the contribution from short distances physics and the result should be added to the low energy contributions.

Numerical estimations have been done by solving the Dyson-Schwinger equation (DSE) \cite{dse} or the Bethe-Salpeter equation (BSE) in the Improved Ladder Approximation (ILA) \cite{bse}. It is interesting that both methods obtain similar results, even if they are very different and do not include the same physics, for example the BSE approach includes resonances while the DSE do not take them into account.
Both calculations use a running coupling constant $\alpha(s)$ \cite{higashijima} that may not be realistic because it does not have the usual long distance properties like Confinement. More will be said below. 
Simpler versions are models like  the Gauged Nonlocal Constituent Quark Model (GNC) \cite{gnc} and the Extended Nambu Jona-Lasinio Models \cite{ENJL} where $S$ and other parameters can be estimated. However they are too simple to predict resonance parameters, for example. The results are shown in table 1.

\begin{center}
\begin{tabular}{|c|c|c|c|c|}\hline
Parameter 			& $S_{\rm QCD}$  	&$-L_{10}\cdot 10^3$ & $F_\pi$ [MeV]  &$S^{\rm WSM}$   \\ \hline
Exper. 			& $0.28\pm 0.035$ \cite{gasser}	& $5.6\pm 0.3\cite{gasser}$ & 94  & $0.49\pm 0.18 $ \cite{experiments} \\ \hline
Free CQ \cite{cqm}			& 0.16	&3.2 & 128 & 0.21\\ \hline
VMD \cite{vmd}  & 0.38	& 7.6&93 & 0.51\\ \hline
$(\rho_V-\rho_A)_{\rm Ex.}$ \cite{sumrules}& 0.33	& 6.6&98 & 0.44 \\ \hline
ACD \cite{acd}	& 0.26 & 5.1 &- & 0.35 \\ \hline
DSE \cite{dse}	& 0.3	& 6 & - &0.4\\ \hline
BSE \cite{bse} 	& 0.43-0.48 & 8.6-9.5& - & 0.24   \\ \hline
GNC \cite{gnc}   &0.24-0.36 &4.7-7.1& 100  &0.32-0.48\\ \hline
ENJL\cite{ENJL} & 0.26-0.31	& 5.1-6.1& 83-89& 0.35-0.41 \\ \hline
\end{tabular}
\end{center}

TABLE 1. Experimental data and Model predictions for  $L_{10}$ parameter (and $S_{\rm QCD}=-16\pi L_{10}$) and $F_\pi$ in the QCD case and for TC models (for $N_{TD}=1$ and $N_{TC}=4$). 
\vskip0.5cm

Given that interactions are strong and nonperturbative contributions may be large one has to use methods like the Dyson-Schwinger equation (DSE) \cite{dse,gnc}, Bethe-Salpeter equation (BSE) \cite{bse} or Lattice QCD. It seems the best method is the BSE in the ILA, besides Lattice that is not considered here. Arguments in favor are that it includes both the Techni-quarks (TQ) self-energy $\Sigma(q^2)$ from DSE and the Vector and Axial vertex form factors $\Gamma^\mu_{V,A}(p,q)$ (including the resonances) from BSE. These two contributions are dominant in the large-N expansion \cite{large-N} over for example pseudogoldstone mesons contributions.
On another side BSE equation is able to explain correctly most of the known properties of QCD like the spectra, the partial decay widths, etc. \cite{lqs-bs}.
 
A very popular TC model assumes $N_{\rm TD}$ techniquark doublets \cite{peskin,N_f} interacting through a new strong force, Technicolor. It is supposed that the TQ are light enough so Chiral Symmetry is preserved. This  is then broken dynamically to produce the Goldstone Bosons, absorbed by the $W$ and $Z$ to adquire their known masses.
A second possibility is to have heavy TQ, so Chiral Symmetry is broken explicitely. In this case the $W$ and $Z$ masses are produced because their self-energies do not vanish at $s=m_W^2<<m_{\rm TQ}^2$ ($\Pi_A(0)\neq 0$ in eq. (\ref{sdef})), even if they do not interact (see eq. (\ref{freeq})). At present nobody knows which possibility is realized by nature, if any. In the case of QCD both, light and heavy quarks produce contributions to $\Pi_A(0)$: for light quarks Chiral Symmetry seems to be broken dynamically, while for the heavy ones is obviously broken explicitely.
The first case was treated for TC within several approximations by authors of refs.  \cite{dse,bse}. Perhaps in that case scaled estimations are a good approximation but the complete calculation has to be done. In the present work it is assumed that the second case is realized in nature so light TQ are absent.
If TQ are heavy the Nonrelativistic Quark Model (NRQM) \cite{nrqm,bs-se,cornell,richardson} is a good approximation like in the case of bottomonium and charmonium physics \cite{richardson}.  It should be emphasized that the light TQ contribution (if any) can not be computed by using the NRQM, due to large relativistic effects so one has to solve the complete BSE.
In the general case of having light and heavy TQ both contributions have to be added.

The purposes of the present work is to review shortlely the calculations \cite{peskin} where QCD is used as an analog computer to show their shortcomings and second to compute the parameter $S$ by using the NRQM for several potentials, constrained by the fact that they have to reproduce the correct electroweak vacuum expectation value $v^2=\sqrt{2}/G_F\simeq (0.245\cdot {\rm TeV})^2 $ (from $ m_W^2=g^2v^2/4$, and $v$ has to be replaced by $F_\pi$ for QCD) and predict a reasonable spectra. First I will work out the simple but physically appealing Cornell potential \cite{cornell} and its particular case, the Coulomb potential. Finally the more realistic Richardson Potential \cite{richardson} is used to get values for $S$ not excluded by experimental data.
 
The parameter $S$ is given by the equations \cite{peskin} 

\begin{eqnarray}
&&\hskip0.3cm S=4\pi \left[ \Pi_{VV}^\prime(0)-\Pi_{AA}^\prime(0)\right]=4\pi\int_0^\infty {{{\rm d}s}\over s}\left[
\rho_V(s)-\rho_A(s)\right]\nonumber \\
&&\hskip0.3cm v^2 = \Pi_{VV}(0)-\Pi_{AA}(0)= \int_0^\infty {\rm d}s\left[\rho_V(s)-\rho_A(s)\right] \label{sdef}
\end{eqnarray}

with $i\left(g^{\mu \nu} \Pi_{VV}(q^2)+B_{VV} q^\mu q^\nu \right)\delta^{ab} \equiv \int {\rm d}^4 x <(J_V(x))^a_\mu (J_V(0))^a_\nu> \exp (iq\cdot x)$ the two point functions, $\rho_V={\rm Im}\left(\Pi_{VV}^\prime (s)\right)/\pi$, and similarly for the axial ones  \cite{sumrules}. The currents are $(J_V)^a_i=\bar{q}T^a\gamma_\mu q$, $(J_A)^a_i=\bar{q}T^a\gamma_\mu\gamma_5 q$ with $\bar{q}=(\bar{u},\bar{d})$  the light quarks doublet for QCD and for TC one has to replace  $\bar{q}$ by $\bar{Q}=(\bar{U},\bar{D})$, the corresponding Tecniquark doublet. Finally,   $T^a$ are the Isospin generators. 
In the case of QCD eqs. (\ref{sdef}) correspond to the \lq zeroth' and first Weinberg Sum Rules \cite{peskin,sumrules} and both are consequences of Chiral Perturbation Theory valid only for the light quarks $u$,  $d$ and $s$  \cite{gasser}. In the case of the Physics Beyond the SM $S$ was already mentionated and the relation for $v$ is obtained from the fact that $W$ and $Z$ masses are generated dynamically due, for example to the presence of new physics like the Higgs or TQ bounded scalars, etc \cite{tc}. Other equivalent definitions in the case of Physics Beyond the SM have been given: 
$S\simeq  h_{AW}\simeq h_{AZ}   \simeq 4s_W^2\epsilon_3/ \alpha $  \cite{peskin}.
The QCD and TC Physics may be different but there are many potential analogies and the formulas are the same, so it is interesting to compare and whenever it is possible do both calculations at the same time.

\section{QCD as an analog computer }

\subsection{Constituent Quarks estimation}

The simplest case is to consider one free constituent quark (or TQ) loop contribution to the two point functions. In the case of TC with $N_{\rm TD}$ degenerate technidoublets and $N_{\rm TC}$ technicolors the result is  \cite{peskin,cqm}

\begin{eqnarray}
S &=& {{N_{\rm TC}N_{\rm TD}}\over{6\pi}}\nonumber \\
v^2 &=& {{N_{\rm TC}N_{\rm TD}}\over{4\pi^2}}m_{\rm TQ}^2\log(\Lambda^2/m_{\rm TQ}^2) \label{freeq}
\end{eqnarray}

For QCD $N_{\rm TD}=1$ and $N_{\rm TC}=3$, and the $L_{10}$ obtained is close to the expected value. On another side $F_\pi$ diverges and a cut-off, $\Lambda$ has to be used. If $m_q=0.3$ GeV and $\Lambda \simeq 1$ GeV are used one obtains the $F_\pi$ shown in table 1. For TC we see that models with $N_{\rm TD} > 4$ are forbidden by experimental data.

The physics here is too simple and interquark forces are obviously negleted. A simple improving is obtained by computing the two loops contribution  \cite{twoloops}, however it is small and in particular no resonances (the main contribution \cite{sumrules} in QCD!) are obtained, an indication that nonperturvative contributions are large. 
It is remarkable, however how this very simple constituent quark model and other ones like GNC \cite{gnc}, ENJL \cite{ENJL}, DSE, BSE, etc., where interactions are somehow taken into account predict values for $L_{10}$ close to experimental data.
It seems that $L_{10}$ does not depend on the interactions neither on the quark masses, at least for QCD  but in general this may not be true. 

\subsection{Scaling VMD results}

In the case of QCD we see from experimental data of fig. 1, for $\rho_V$ and $\rho_A$ that the main contributions to the sum rules of eq. (\ref{sdef}) are from the two lightest resonances: $\rho(770)$ and $a_1(1260)$. 
On another side it is found that the contributions of the resonance widths  
are very small and a good approximation is to take $\rho_V =F_\rho^2\delta \left( s-m_\rho^2\right)$ and $\rho_A=F_A^2\delta \left( s-m_A^2\right)$, this is the so called narrow resonance approximation. In this case, for two resonances one has

\begin{eqnarray}
S=4\pi \left[ {{F_\rho^2}\over{m_\rho^2}}-{{F_A^2}\over{m_A^2}}\right] \hskip0.4cm v^2 \to F_\pi^2=F_\rho^2-F_A^2 \label{resonances}
\end{eqnarray} 
 
The usual approach \cite{peskin} is to estimate $S$  for TC theories (at least for QCD-like ones) \cite{peskin} by using the following relations: 1) the second Weinberg sum rule $m_\rho^2 F_\rho^2=m_A^2 F_A^2$   \cite{peskin,sumrules}, 2) the KSRF relation $m_{\rho_T}^2/m_{A_{T}}^2=m_\rho^2/m_A^2=1/2$ \cite{ksrf}  and 3) the scaling ($N_C=3 \to N_{TC}$) relation $F_{\rm TC}^2/m_{\rho_{\rm TC}}^2= N_{\rm TD}(N_{\rm TC}/3)F_\pi^2/m_\rho^2$ \cite{scaling,large-N}.
Notice that the first two relations are approximately satisfied in the QCD case.
Using the first and second Weinberg sum rules and the KSRF relation one obtains $S=6\pi F_\pi^2/m_\rho^2$ (or $L_{10}=-3F_\pi^2/8m_\rho^2$) and  after scaling  

\begin{eqnarray}
S=6\pi {{F_\pi^2}\over{m_\rho^2}} \to 6\pi N_{TD}{{N_C}\over 3} {{F_\pi^2}\over{m_\rho^2}}=N_{TD}{N_{TC}\over 3} S_{\rm QCD}
\end{eqnarray}

To evaluate $S_{\rm QCD}$ one has several alternatives: compute it directly form $F_\pi$ and $m_\rho$, so $S_{\rm QCD}\simeq 0.28$. An intermediate possibility is to use several of the relations mentioned above to get values for the constants that may not be very well known experimentally, like $F_A$ and so on.
Another possibility is to use the experimental values \cite{pdg} in eq. (\ref{resonances}): $F_V=154\pm 3$ MeV  obtained from $\Gamma (V\to l^- l^+)= 4\pi\alpha^2 F_V^2/3 m_V\simeq (6.9\pm 0.3)\hskip0.2cm{\rm KeV}$ 
and $F_A=123 \pm 25$ MeV from $\Gamma (a_1\to \pi \gamma)=  (\alpha F_A^2m_a/24F_\pi^2)\left(1-m_\pi^2/m_a^2\right)^3\simeq (0.64\pm 0.25)\hskip0.2cm{\rm  MeV}$. 
Using these values one obtains $S_{\rm QCD}=0.38$ ($L_{10}=-0.0076$) and $F_\pi=93$ MeV in agreement with the QCD experimental results. 
Finally one can use directly the measured spectral functions $\rho_V(s)-\rho_A(s)$ (see fig. 1) \cite{peskin,sumrules}, to obtain $S_{\rm QCD}=0.33$ and $F_\pi=98$ MeV in agreement with the expectations. 
Alternative approaches are the  VMD in the Lagrangian version \cite{vmd} but the physics is the same. In any case the value for $S_{\rm QCD}$ are not very different from the one obtained in the free constituent TQ case and again TC models with $N_{\rm TD}> 4$ are excluded.

\begin{figure}[b!] 
\centerline{\epsfig{figure=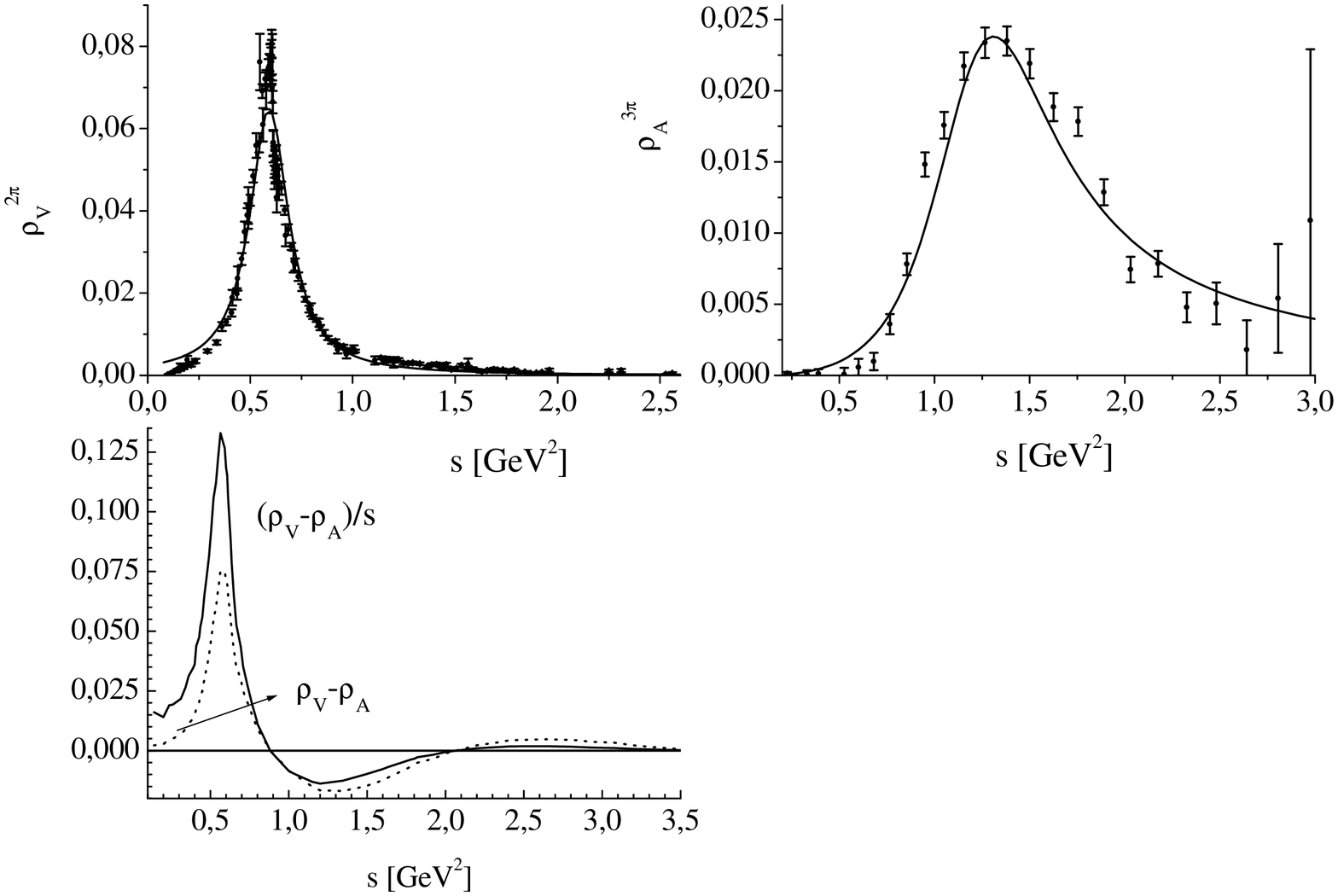,height=8cm,width=20cm}}
\vspace{10pt}
 \label{fig1}
\end{figure}
\vskip-0.5cm
{FIG. 1.} Experimental values for $\rho_{V,A}$ and the \lq best fit' line for the integrands of the sum rules in eq. (\ref{sdef}), from ref. \cite{sumrules}.
\vskip0.5cm

\subsection{Critics}

As mentioned before there are several critics to this approach, let's see. The first objection is to the scaling hypothesis: this relation is only the leading contribution, in the large-N expansion \cite{large-N} and we do not know if it is valid for a general theory, specially when Strong Interactions and their nonlinear character may produce large nonperturbative contributions. In any case an explanation is due to the fact that the result is independent of the interactions for a general theory.
A second objection is to the KSRF relation that in the case of heavy quarks is clearly wrong because the hyperfine splitting is certainly smaller. The second Weinberg sum rule may be wrong in the general case too. 
Another difficulty is the fact that the QCD Sum Rules are saturated by only two resonances, the $\rho(770)$ and $a_1(1260)$ so heavier resonances were not taken into account to get $S$. However this may not be the case in TC Models, even in QCD-like ones and even less in WTC models (For Walking Technicolor \cite{wtc,N_f,critics} the coupling constant \lq walks' as $\alpha_{\rm TC}(s) \sim $ constant for $\Lambda_{TC} \sim 1 \hskip0.15cm{\rm TeV} <\sqrt{s}<\Lambda_{\rm ETC}$ so Asymptotic freedom is retarded and high energy physics, like heavy resonances may produce large contributions to low energy parameters).

According to the scaled results heavy TQ do not decouple from the Standard Model Physics \cite{decoupling}. This is not the case in QCD where for example hadronic physics is completely blind to the top quark physics.
One can imagine that heavy TQ form new $W$-s and $Z$-s ($W^*$, $Z^*$, etc), but they do not contribute to the masses of the known ones. This is the case in QCD where for example $b$ quarks form $\Upsilon$ resonances but they do not contribute to the mass of the $\rho(770)$ for example. It may be that TQ doublets are grouped in \lq light' and \lq heavy' ones, and their contributions decreases when their masses become large.

On another side calculations assume Isospin is a good symmetry \cite{isospin}. One can argues that in general this could not be the case so doublets may not be degenerate. For the known quarks and leptons doublets non degeneracy is the rule and only the first quark doublet is approximately degenerate!. Isospin seems to be present due to two accidents: the smallness of the fine structure constant and the approximate degeneracy of $u$ and $d$ quarks, thus $\rho \sim 1$  even if  $m_t >> m_b$. In order to take into account these facts Extended Technicolor theories \cite{etc} assume that Physics present at energies of the order $\Lambda_{ETC}$ should conserve Isospin to keep $m_u\sim m_d$ and $\rho \sim 1$, and the Physics at energies of the order of $\Lambda_{TC}$ should break it to explain the top quark mass. Taking this in mind $S$ is computed and a negative value can be obtained \cite{negative-S}. One should be aware too, that TQ may not belong to doublets but to higher dimensional representations having in general large mass differences.

Finally, it was assumed that {\em all} TQ doublets are degenerate without having any reason for that. This is a very strong hypothesis, specially when in the case of known quarks and leptons  degeneracy is not the rule but on the contrary, there are large mass differences between for example neutrinos and top quark.

\section{Nonrelativistic Quark Model}

As mentioned before the main purpose of this paper is to work out the NRQM prediction for $S$ by  modeling the TQ-TQ interaction with a potential, obtained by extrapolating the QCD one.
Before going to the calculation I want to argue in favor of this approach. If the TQ-s are heavy enough the relativistic effects are very small, the coupling constant becomes small due to Asymptotic freedom and the TQ Self-Energies become constant and equal to the TQ masses, $\Sigma(q^2) \simeq m_{\rm TQ}$. Finally it has been found that using relativistic equations do not improve necessarily the results because the parameters can be redefined \cite{relativistic}. In order to compute $S$ and $v^2$ one can use eqs. (\ref{resonances}), once  the contributions from all the resonances are considered.
According to the NRQM the $F_V$ and $F_A$ are given by \cite{nrqm,peskin}.

\begin{eqnarray}
F_V^2 = {{N_{\rm TC}}\over{2\pi}}{{|R_{ns}(0)|^2}\over{m_V}}  \hskip0.5cm
F_A^2 ={{24N_{\rm TC}}\over \pi}{{|R_{np}^\prime(0)|^2}\over{m_a^3}}
\end{eqnarray}

for each doublet and $R(r)$ satisfy the radial Schr\"odinger Equation

\begin{eqnarray}
-{1\over{2\mu}}\left[{{{\rm d}^2}\over{{\rm d}r^2}}+{2\over r}
{{\rm d}\over{{\rm d}r}}\right]
{\rm R}(r)+\left[V(r)+{{l(l+1)}\over{2\mu r^2}}\right]{\rm R}(r)=E{\rm R}(r) \nonumber \\
\end{eqnarray}

with $\mu$ the reduced mass and ${\rm R}(r)$ satisfy the usual boundary and  normalization conditions.

\subsection{Cornell and Coulomb Potentials}

The Cornell Potential \cite{cornell} is $V=2m_{TQ}(d-1)+Fr-a/r$, with $a=C_2(R)\alpha_s$,  $C_2(R)=(N^2-1)/2N$ and $F$ is the \lq String tension'. The first term is some kind of background not considered in the literature that may be turned off by taking $d=1$. Redefining ${\rm R}=(m_{\rm TQ}F)^{1/2}u/r$ and $x=(2\mu F)^{1/3}r$ we have

\begin{eqnarray}
-u^{\prime \prime}+V_{\rm eff.}u \equiv -u^{\prime \prime}+\left[x-b/x+l(l+1)/x^2\right]u=\epsilon_{nl} u \label{se}
\end{eqnarray}

with $u(0)=u(\infty)=0$, $u(x)$ normalized to unity and $b=a((2\mu)^2/F)^{1/3}$. The wavefunction and its first derivative at the origin are $|{\rm R}_{ns}(0)|^2 =m_{\rm TQ} F |u_{ns}^\prime(0)|^2$ and $|{\rm R}_{np}^\prime(0)|^2 =(m_{\rm TQ} F)^{5/3}|u^{\prime \prime}_{ np}(0)|^2/4$, respectively.
Notice that $\epsilon_{nl}$, $u^\prime_{nl}(0)$ and $u^{\prime\prime}_{nl}(0)$ depend only on the parameter $b$ and can be obtained by solving eq. (\ref{se}) numerically \cite{cornell}.   
The masses of the vector ($l=0$), the axial vector ($l=1$) resonances and the hyperfine splitting are 

\begin{eqnarray}
&&m_{nl} = 2m_{\rm TQ}d+{{(m_{\rm TQ} F)^{2/3}}\over {m_{\rm TQ}}}\epsilon_{nl} \nonumber \\
&&m(n^3{\rm s}_1) -m(n^1{\rm s}_0)={{2(m_{\rm TQ}F)^{4/3}}\over{3m_{\rm TQ}^3}}
b|u^\prime_{ns}(0)|^2
\end{eqnarray}

where the second relation was obtained from the spin-spin potential $V_{s-s}=8\pi C_F\alpha_S(\vec{\bf s_1}\cdot \vec{\bf s_2})\delta^{(3)}(\vec{\bf r})/3m_{\rm TQ}^2$ \cite{nrqm}.

For $N_{\rm TD}$ degenerate doublets we have 

\begin{eqnarray}
S &=& N_{\rm TD}{{N_{\rm TC}}\over 4}t^{3/2}\sum_n
\left[{{|u_{ns}^\prime(0)|^2}\over{(d+t\epsilon_{ns}/2)^3}}-3t{{|u_{np}^{\prime\prime}(0)|^2}\over{(d+t\epsilon_{ns}/2)^5}}\right] \nonumber \\
v^2 &\equiv& {{\sqrt{2}}\over{G_F}}= N_{\rm TD}{{N_{\rm TC}}\over{4\pi}}m_{\rm TQ}^2t^{3/2}\sum_n
\left[{{|u_{ns}^\prime(0)|^2}\over{d+t\epsilon_{ns}/2}}-3t{{|u_{np}^{\prime\prime}(0)|^2}\over{(d+t\epsilon_{ns}/2)^3}}\right]\label{scornell}
\end{eqnarray}

where $t=(F/m_{\rm TQ}^2)^{2/3}$.

\begin{figure}[b!] 
\centerline{\epsfig{file=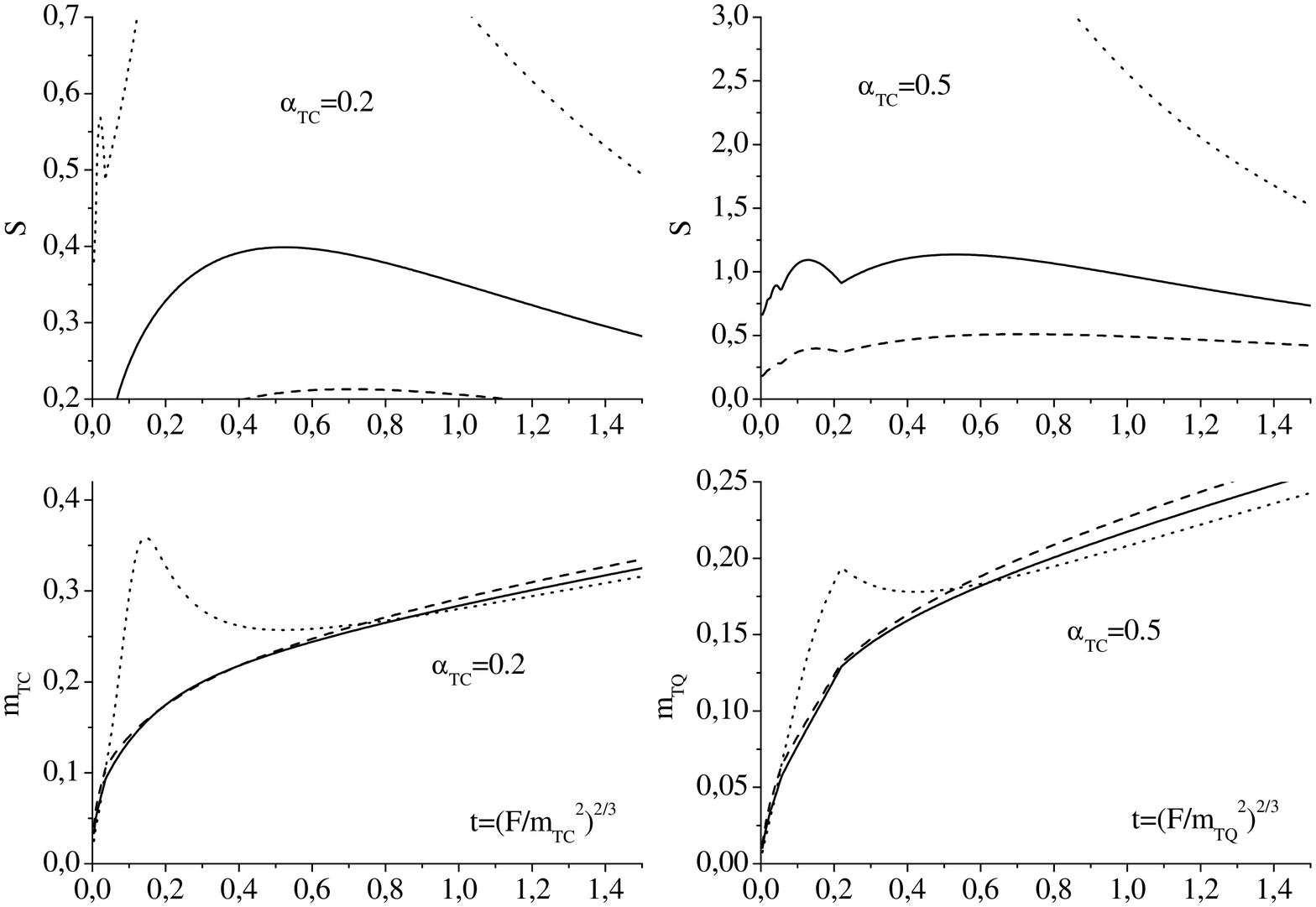,height=10cm,width=16cm}}
\vspace{10pt}
\label{fig2}
\end{figure}

\vskip-0.5cm
FIG. 2. Values for $S$ and  $v^2$, for $N_{\rm TC}=4$, $N_{\rm TD}=1$. The dotted, continuous and dashed lines are for $d=0.5,1$ and 1.5, respectively. In the case of $\alpha_{\rm TC}=0.2$, following the explanation given in the text for $S\simeq 0.4$ and $d=1$ we get $t\simeq 0.5$ and from it $m_{\rm TQ}\simeq 0.22$ TeV and $F\simeq 0.017$ TeV$^2$. The mass for the Techni-$\rho$  is then 0.64 TeV and the hyperfine splitting 0.04 TeV, so relativistic corrections are small and the NRQM is valid for these values. Notice that for $d=1.5$ one obtains $S\sim 0$ for all the values of $t$. The case of $\alpha_{\rm TC}=0.5$ and $d=1$ may be rule out because $S\sim 0.9$ for the whole intervale. Light TQ contributions were not considered.
\vskip0.5cm

In order to analyze eq. (\ref{scornell}) it is reasonable to keep $N=N_{\rm TC}=4$, $N_{\rm TD}=1$, $\alpha_{\rm TC}$ and $d$ fixed so $S$ and $v^2$ depend only on $t$ ($b=a/\sqrt{t}$).
Fig. 2 illustrate the behavior of $S$ and $m_{\rm TQ}$, for two values of $\alpha_{\rm TC}$: for a given value of $S$ one obtains the corresponding $t$, then from the second graph the $m_{\rm TQ}$ needed to have the correct $v^2$ and the \lq String tension', $F$ can be found. From them the spectra can be computed.
A good place to test these predictions is of course in the case of light quarks in QCD, but unfortunately for light quarks large relativistic correction are present and NRQM is not a good approximation. If one ignore this difficulty a value of $S\sim 0.28$ is obtained, that is no far from the experimental result. 

One important limit of the Cornell potential is the very heavy TQ case. There one can see that $b$ grows like $m_Q^{2/3}$ and the potential becomes coulombic at short distances, independent on the confinement phenomena. Now it is easy to obtain analytic solutions for the spectra: $m_{nl}=2m_{\rm TQ}[d-(a/n)^2/8]$, for the wave function $|R_{ns}(0)|^2=4(a\mu/n)^3$ and for its first derivative $|R_{ns}^\prime(0)|^2=4(n^2-1)(a\mu/n)^5/9$. From these expresions $S$ and $v^2$ are

\begin{eqnarray}
S &=& N_{\rm TD}{{N_{\rm TC}}\over 8}\sum_n\left[{{a/n}\over{d-(a/n)^2/8}}\right]^3
\left[1-{1\over 3}{{(a/n)^2(n^2-1)}\over{[d-(a/n)^2/8]^2}}\right]
\nonumber \\
v^2 &=& N_{\rm TD}{{N_{\rm TC}}\over{8\pi}}m_{\rm TQ}^2\sum_n{{(a/n)^3}\over{d-(a/n)^2/8}}
\left[1-{1\over 3}{{(a/n)^2(n^2-1)}\over{[d-(a/n)^2/8]^2}}\right] 
\end{eqnarray}

The behavior of $S$ and of the TQ mass, needed to have the correct $v$ are shown in Fig. 3 for the case of the Coulomb potential. Notice that at large distances the confining potential is still present and no continuum contribution is obtained.

\begin{figure}[b!] 
\centerline{\epsfig{file=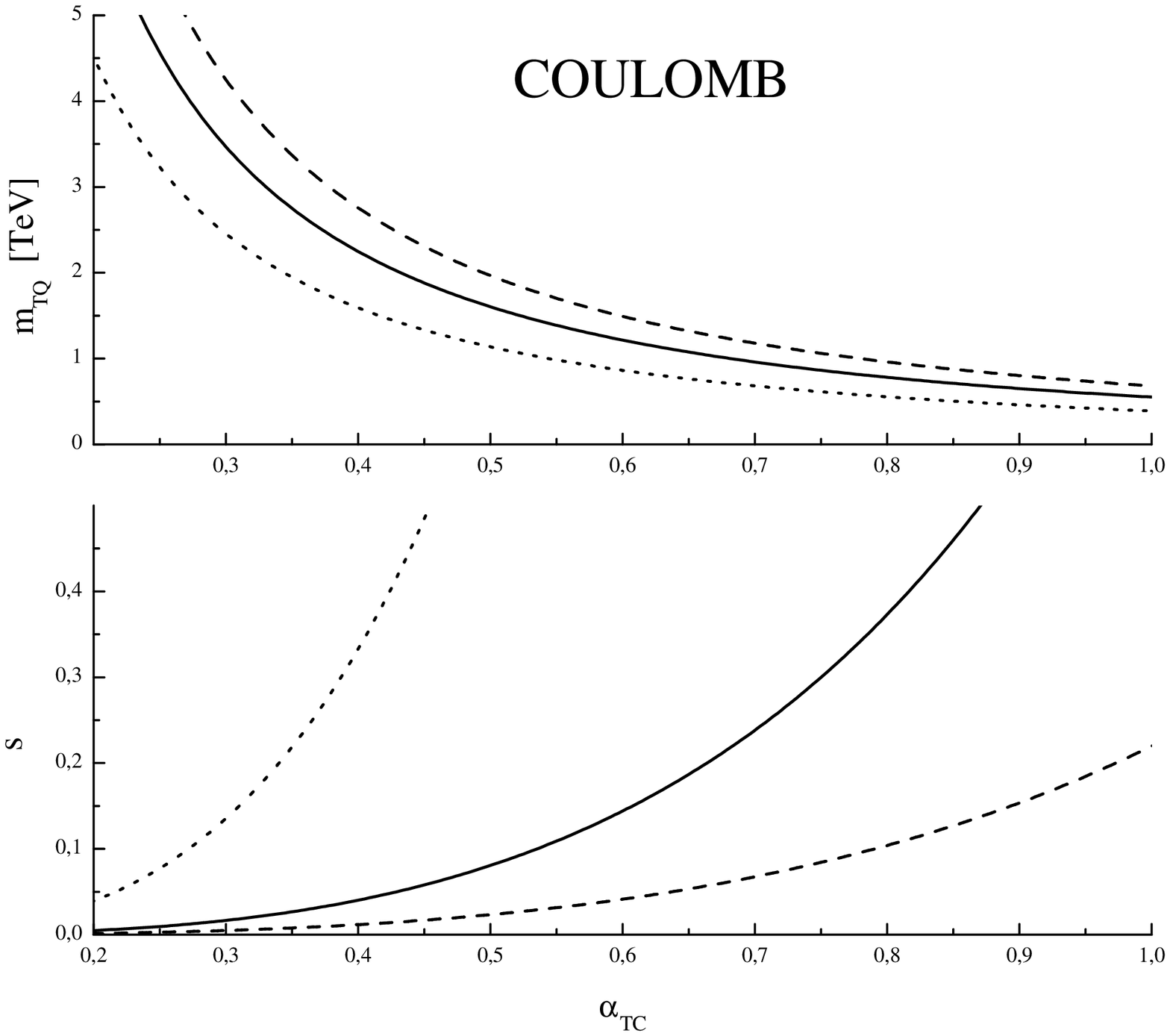,height=8cm,width=16cm}}
\vspace{10pt}
\label{fig3}
\end{figure}

\vskip-0.5cm
FIG. 3. Values for $S$ and  $v^2$ for the Coulomb case, as in fig. 2. We see that $S\to 0$ when $m_{\rm TQ}\to \infty$ (so $\alpha_{\rm TC}(m_{\rm TQ})\to 0$).
\vskip0.5cm

In the limit of heavy TQ it is easy to see that $m_{V,A} \to m_Q$,  $R(0)^2 \to \alpha_S^3(m_Q)m_Q^3$ and $F_V^2 \to \alpha_S^3(m_Q)m_Q^2$, so one obtains finally that $S\to \alpha_S^3(m_Q) \to 0$ (see fig. 3) and  very heavy quarks decouple, in agreement with  Appelquist-Carazzone theorem \cite{decoupling} and contradicting scaling assumptions.  
However the Coulomb potential is a simple one bare gluon exchange  approximation that has to be corrected even at very high energies  \cite{richardson}. 

\subsection{Richardson Potential}

As mentioned before the calculations done by now, using the DSE and BSE equations use a running coupling constant $\alpha(s)$ that may not be realistic because it does not have the usual long distance confinement properties \cite{higashijima}, thus in order to break Chiral Symmetry dynamically \cite{csb}  a very strong potential is needed.  Besides the $\alpha(s)$ is in disagreement with lattice calculations for QCD \cite{latt-pot}. A more realistic potential  is the so called Richardson potential \cite{richardson}. In general the potential is given by 

\begin{eqnarray}
V(r)&=& \int {{{\rm d}^3q}\over{(2\pi)^3}}{\rm e}^{iq\cdot r}V(q) ={1\over{2\pi^2 r}}\int_0^\infty V(q^2)q\sin(qr){\rm d}q 
\end{eqnarray}

with $V(q^2) =-C_2(R)4\pi \alpha(q^2)/q^2$. The Richardson potential is obtained by extrapolating the Asymptotic freedom expression to the infrared region as $\alpha^{\rm AF}(q^2) = 4\pi/\beta_0\log\left(q^2/\Lambda^2 \right)\to \alpha^{\rm Rich.}(q^2) = 4\pi/\beta_0\log\left(1+q^2/\Lambda^2 \right)$  \cite{richardson}, with $\beta_0=11C_2(G)/3-2n_F/3$, $C_2(G)=N$ and $C_2(R)=(N^2-1)/2N$. Then one has for the Richardson potential

\begin{eqnarray}
V_{\rm Rich.}(r) & =& -{{8C_2(R)}\over{\beta_0 r}}\int_0^\infty {1\over{\log\left(1+q^2/\Lambda^2\right)}}{{\sin(qr)}\over q}{\rm d}q \nonumber \\
&=& Fr-{{2\pi C_2(R)f (\Lambda r)}\over{\beta_0 r}}  \to \cases{ -2\pi C_2(R)/3\beta_0 r\log(1/\Lambda r)  ,&if $r \to 0$\cr
	Fr & if $r\to \infty $\cr} 
\end{eqnarray}

with the \lq String tension', $F=2\pi C_2(R)\Lambda^2/\beta_0$ and

\begin{eqnarray}
f(x) = {4\over \pi}\int_0^\infty {\rm d}k{{\sin(kx)}\over k}\left[{1\over{\log(1+k^2)}}-{1\over{k^2}}\right]=1-4\int_1^\infty {{{\rm d}k}\over k}{{\exp(-kx)}\over{\pi^2+[\log(k^2-1)]^2}}
\end{eqnarray}

The linear part explains confinement as well as the Regge trajectories \cite{regge} and it is able to produces the Dynamical Chiral Symmetry breaking \cite{csb}. Besides Richardson potential agrees with lattice calculation \cite{latt-pot} as is shown in fig. 4. Finally it is able to explain correctly the heavy quarks spectra and  their leptonic widths \cite{richardson}.

\begin{figure}[b!] 
\centerline{\epsfig{file=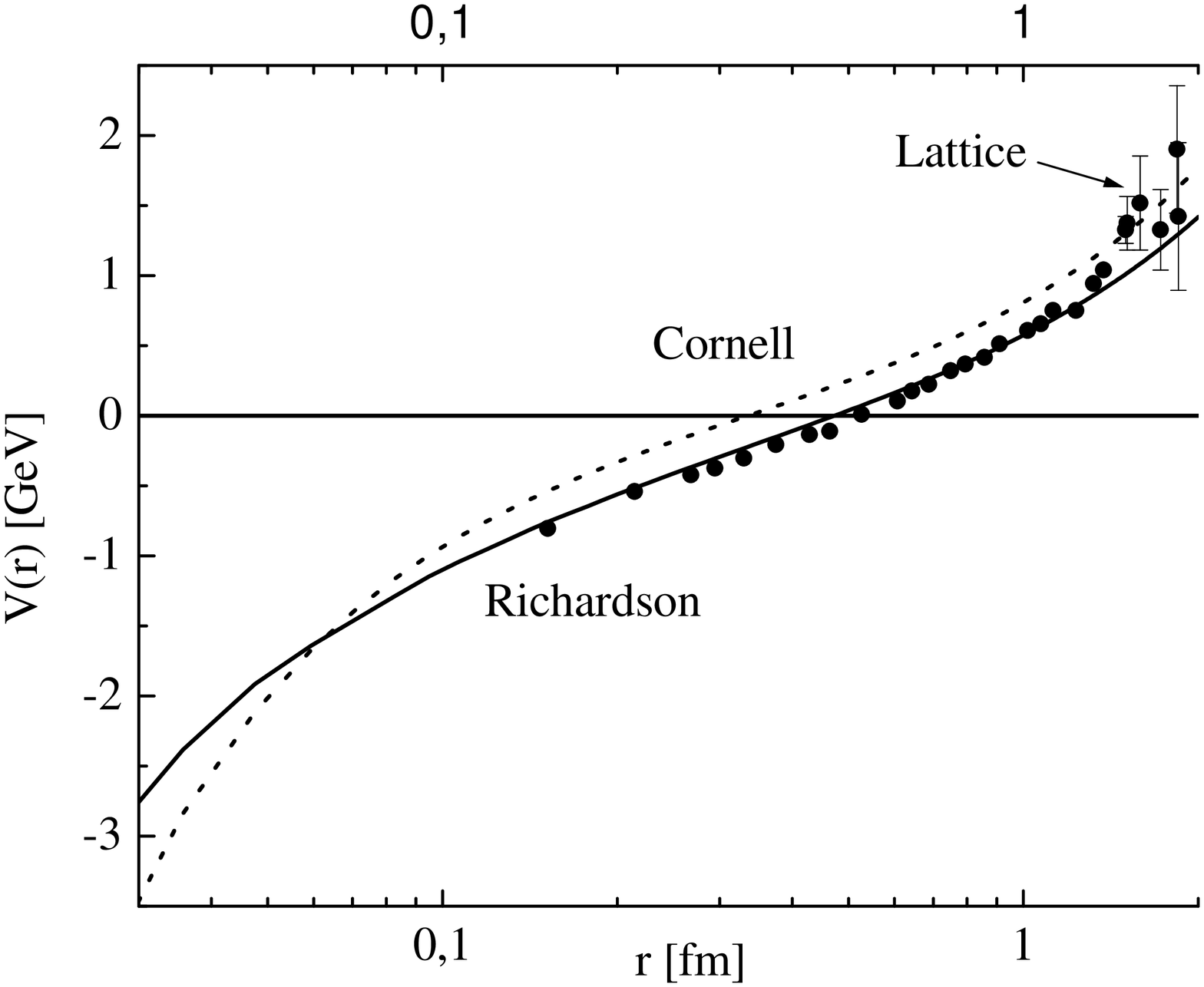,height=6cm,width=14cm}}
\vspace{10pt}
\label{fig4}
\end{figure}
\vskip-0.5cm
FIG. 4. Cornell (dashed line) \cite{cornell} and Richardson (continuous one) \cite{richardson} potentials compared with  lattice calculations of the CP-PACS group \cite{latt-pot} (dots), for QCD.  
\vskip0.5cm

Transforming the wavefunction as $R(r)=\Lambda^{3/2}u(x)/x$, with $x\equiv \Lambda r$ one obtains eq. (\ref{se}) for $u(x)$ with $V_{\rm eff.}(x)=(m_{\rm TQ}/\Lambda)V_{\rm Rich.}/\Lambda +l(l+1)/x^2$. One important property of eq. (\ref{se}), in the case of the Richardson potential is that it has only one input parameter, $m_{\rm TQ}/\Lambda$. The boundary and normalization conditions for $u$ are the same of the Cornell potential case.
The masses of the vectorial resonances are then $m_{nl}=2m_{\rm TQ}+E_{nl}=2m_{\rm TQ}+\Lambda^2\epsilon_{nl}/m_{\rm TQ}$, where $\epsilon_{nl}$ are obtained from (\ref{se}). The wavefunction and its first derivative are $|R_{ns}(0)|^2=\Lambda^3 |u_{ns}^\prime(0)|^2$, $|R_{np}^\prime(0)|^2=\Lambda^5 |u_{np}^{\prime\prime}(0)|^2/4$, respectively. Finally, the hyperfine splitting and the decay constants are  

\begin{eqnarray}
& &m(n^3{\rm s}_1)-m(n^1{\rm s}_0)={{2C_2(R)\alpha((2m_{\rm TQ})^2)\Lambda^3}\over{3m_{\rm TC}^2}}|u_{ns}^\prime(0)|^2 \nonumber \\
& &F_V^2 = {{N_C\Lambda^3}\over{2\pi m_V}}|u_{ns}^\prime(0)|^2 \hskip0.5cm
F_A^2 = {{6N_C\Lambda^5}\over{\pi m_a^3}}|u_{np}^{\prime\prime}(0)|^2
\end{eqnarray}

Notice that $\epsilon_{nl}$, $u_{ns}^\prime(0)$, $u_{np}^{\prime \prime}(0)$ and therefore $S$ and $(v/\Lambda)^2$ depend only on $m_{\rm TQ}/\Lambda$. The Sch\"odinger equation (\ref{se}) was solved numerically and the results are given in tables 2 and 3 for two specific values of $m_{\rm TQ}/\Lambda$.
In order to get $S$ and $v$ the first 10 resonances were taken into account assuming that the series converge rapidly due to the fact that for example the widths of the heavier resonances are large enough so their contributions can be negleted. This is the case in QCD where only the first two resonances  contribute to the sum rules. Notice that for the Cornell and Richardson potentials there is no continum part in the spectra because there are no free quarks (or TQ).
From tables like 2 and 3, the corresponding value for $(v/\Lambda)^2$ and again  constraining $v$ to its experimental value one obtains $\Lambda$. Now, from the input parameter $m_{\rm TQ}/\Lambda$ one obtains $m_{\rm TQ}$ and from it the respective spectra can be calulated. 

\begin{center}
\begin{tabular}{|c|c|c|c|c|c|c|c|c|}\hline
n& $\epsilon_{ns}$ & $|u_{ns}^\prime(0)|^2$ & $\epsilon_{np}$ & $|u_{np}^{\prime \prime}(0)|^2$   &$m_V/\Lambda$ & $(F_V/\Lambda)^2$ &$m_a/\Lambda$ & $(F_a/\Lambda)^2$ \\ \hline
 1& 1.4639(7)& 1.6927& 2.6594(8)&	1.5202&3.46 &0.311&4.66 &0.115 \\ \hline
 2& 3.3048(9)& 1.3590& 4.1548(9)&	2.3018&5.3  &0.163&6.15 &0.076 \\ \hline
 3& 4.7168(10)&1.2514& 5.4252(11)&	2.8966&6.72 &0.119&7.43 &0.054 \\ \hline
 4& 5.9373(12)&1.1973& 6.5658(13)&	3.3894&7.94 &0.096&8.57 &0.041 \\ \hline
 5& 7.0465(13)&1.1732& 7.6256(13.5)&3.8734&9.05 &0.083&9.63 &0.033 \\ \hline
 6& 8.0823(14)&1.1497& 8.6169(15)&	4.1787&10.08&0.073&10.62&0.027 \\ \hline
 7& 9.0439(15)&1.1108& 9.5419(16)&	4.4944&11.04&0.064&11.54&0.022 \\ \hline
 8& 9.9533(16)&1.1040&10.4275(16.5)&4.7889&11.95&0.059&12.43&0.019 \\ \hline
 9&10.8208(17)&1.0829&11.2727(17.5)&5.0570&12.82&0.054&13.27&0.016 \\ \hline
10&11.6548(18)&1.0743&12.0878(18.5)&5.2437&13.65&0.05 &14.09&0.014 \\ \hline
\end{tabular}
\end{center}

TABLE 2. Energy eigenvalues and wavefunctions at the origin for the Richardson potential. The input parameters are $m_{\rm TQ}/\Lambda=1.0$, $N_{\rm TC}=4$ and $N_{\rm TD}=1$. The results are $S\simeq 0.37$ and $(v/\Lambda)^2\simeq 0.7$ so $\Lambda\sim 0.3$ TeV, $m_{\rm TQ}\sim 0.3$ TeV and the lightest vectorial resonance mass is $m_{\rho_{TC}}=1$ TeV. Besides $\Delta m_{H. F.} \simeq 0.38$ TeV and  $\alpha (2m_{\rm TQ})\sim 0.6$ so relativistic effects are small.
\vskip0.5cm

\begin{center}
\begin{tabular}{|c|c|c|c|c|c|c|c|c|}\hline
n& $\epsilon_{ns}$ & $|u_{ns}^\prime(0)|^2$ & $\epsilon_{np}$ & $|u_{np}^{\prime \prime}(0)|^2$   &$m_V/\Lambda$ & $(F_V/\Lambda)^2$ &$m_a/\Lambda$ & $(F_a/\Lambda)^2$ \\ \hline
 1&-48.61(1.5)&552.85&-11.78(2)  &5843.74&58.38&6.029&59.61&0.211 \\ \hline 
 2&-4.56(2.1)&206.93 &12.25(2.4) &5399.83&59.85&2.201&60.41&0.187 \\ \hline 
 3&17.69(2.7)&139.92 &29.33(3)   &5042.46&60.59&1.47&60.98&0.17 \\ \hline 
 4&33.99(3.2)&112.97 &43.31(3.4) &4859.84&61.13&1.176&61.44&0.16 \\ \hline 
 5&47.50(3.8)&98.36  &55.49(4)   &4672.26&61.58&1.017&61.85&0.151 \\ \hline 
 6&59.38(4)&  89.12  &66.50(4.2) &4625.49&61.98&0.915&62.22&0.147 \\ \hline 
 7&70.16(4.6)&82.57  &76.65(4.8) &4457.64&62.34&0.843&62.56&0.139 \\ \hline 
 8&80.14(4.8)&77.79  &86.16(5)   &4419.81&62.67&0.79&62.87&0.136 \\ \hline 
 9&89.50(5.2)&73.98  &95.15(5.4) &4305.79&62.98&0.748&63.17&0.13 \\ \hline 
10&98.37(5.4)&70.94  &103.71(5.6)&4260.79&63.28&0.714&63.46&0.127 \\ \hline 
\end{tabular}
\end{center}

TABLE 3. Same as in table 2 but now the input parameters are $m_{\rm TQ}/\Lambda=30.0$ $N_{\rm TC}=4$ and $N_{\rm TD}=1$.  The results are $S\simeq 0.05$ and $(v/\Lambda)^2\simeq 14.3$ so $\Lambda\sim 0.066$ TeV, $m_{\rm TQ}\sim 1.98$ TeV  and $m_{\rho_{TC}}=3.85$ TeV. Besides $\Delta m_{H. F.} \simeq 0.006$ TeV and $\alpha (2m_{\rm TQ})\sim 0.11$ so relativistic effects are very small.
\vskip0.5cm

The dependence of the relevant parameters on the TQ mass, $m_{\rm TQ}/\Lambda$ is shown in table 4. and fig. 5. One can see how very heavy TQ-s decouple and do not contribute to $S$.

\begin{center}
\begin{tabular}{|c|c|c|c|c|}\hline
$m_{\rm TQ}/\Lambda$& $S$ & $(v/\Lambda)^2$ &  $(F_V/\Lambda)^2$& $(F_a/\Lambda)^2$ \\ \hline
1 & 0.37& 0.7&0.311&0.115\\ \hline
5 & 0.235&2.4&1.2&0.27 \\ \hline
10&0.15 &5.2&2.11&0.236 \\ \hline
20&0.076&9.96&3.97&0.212 \\ \hline
30&0.05 &14.3&6.03&0.211 \\ \hline
40&0.037&18.6&8.25&0.216 \\ \hline
50&0.029 &22.9&10.6&0.22 \\ \hline
\end{tabular}
\end{center}

TABLE 4. $S$, $(v/\Lambda)^2$, etc., for several values of the TQ mass,  $m_{\rm TQ}/\Lambda$.

\begin{figure}[b!] 
\centerline{\epsfig{file=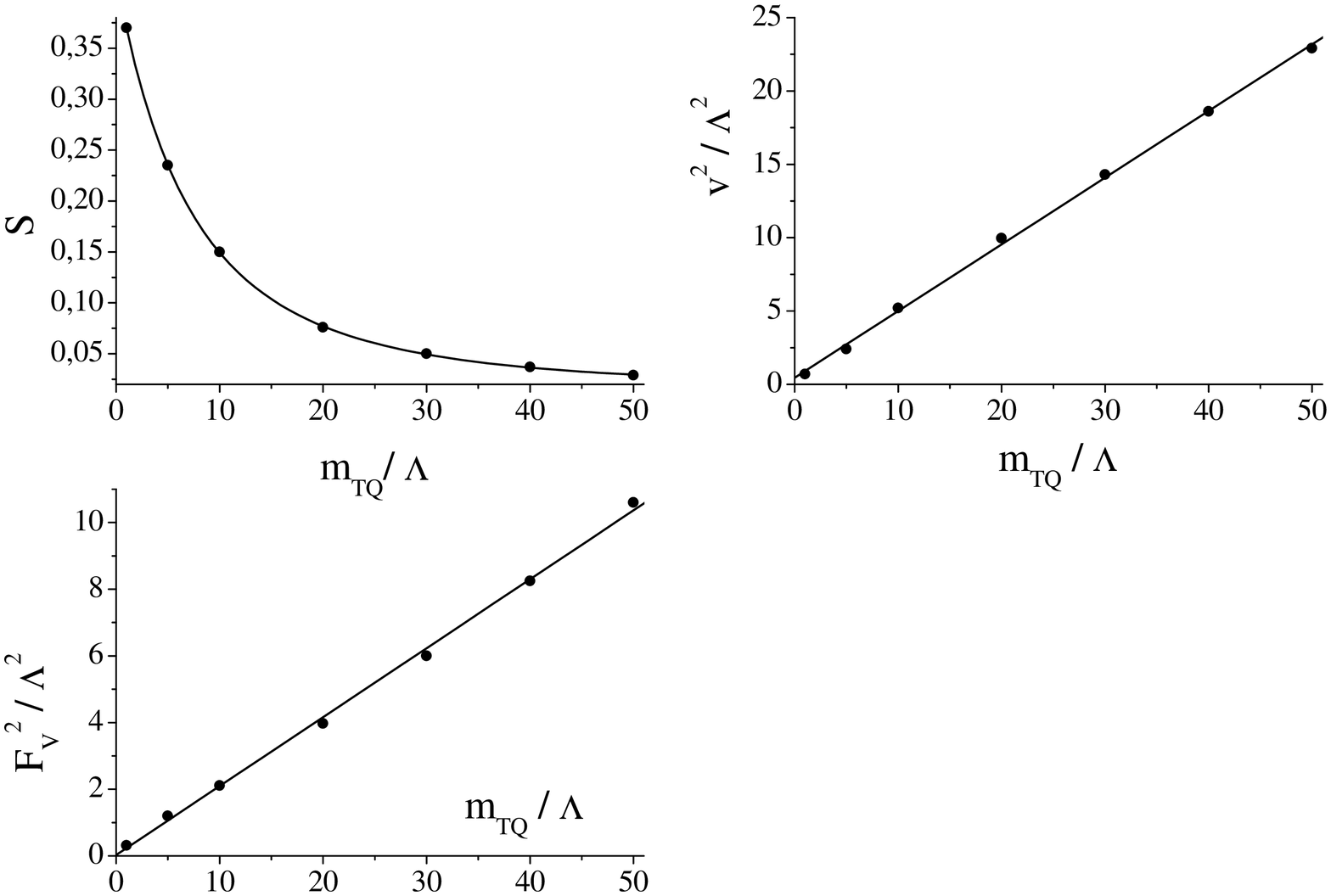,height=8cm,width=18cm}}
\vspace{10pt}
\label{fig5}
\end{figure}

\vskip-0.5cm
FIG. 5. $S$ and $v^2/\Lambda^2$ for the Richardson Potential. The points are fitted by the formula $S=0.016+0.4/(1+0.05\cdot(m_{\rm TQ}/\Lambda))^{2.63}$, decreasing with $m_{\rm TQ}$ and $v^2/\Lambda^2= 0.45+0.45 \cdot(m_{\rm TQ}/\Lambda)$. Finally $F_V^2/\Lambda^2= 0.02+0.21\cdot(m_{\rm TQ}/\Lambda)$. Light TQ contributions were not considered.
\vskip0.5cm

\section{Conclusions}

I found, using the NRQM that it is possible to produce values for $S$ consistent with experimental 
constrains for simple QCD-like TC models. The vacuum expectation value $v$ was kept equal to its experimental value.
Two potentials were used: the Cornell one and the more realistic Richardson potential. 

It was obtained that $S$ decrease with $m_{\rm TQ}$ like $\alpha^3(m_{\rm TQ})$ in the case of the Cornell Potential and approximately like $1/m_{\rm TQ}^{2.63}$ for the Richardson Potential. This  contradicts scaling results \cite{peskin} where no dependence was obtained and  agrees with decoupling theorem \cite{decoupling}. In this way we can obtain simple TC Models that are not rule out by experimental constrains on $S$. For more general models (WTC) more parameters are involved and this situation should be reproduced without major problems.
For $v$, and for the Leptonic decay constants, $F_V$ and $F_A$ it is found that in the NRQM limit both grow for the Richardson potential as $\sqrt{m_{\rm TQ}}$, a softer dependence that the case of the free quark, where they grow like $m_{\rm TQ}$ and in the case of the Cornell potential where they growing is intermediate between these two cases.
The result for $S$ is more solid because it is less sensitive to high energy physics and the sum rule used to compute it converges rapidly. Finally, it was found that $S$ depends on the Interactions and on the TQ masses, as expected.

\acknowledgments

I want to thank to J. Donoghue for the original idea on which this article was developed. This research was supported in part by Colciencias (Colombia).


\begin{references}
\bibitem{peskin}  M. Peskin, T. Takeuchi, Phy Rev. Lett. {\bf 65}, 964 (1990); Phys. Rev. {\bf D46}, 381 (1992). \\
D. C. Kennedy, B. W. Lynn, Nucl. Phys. {\bf B 322},1 (1989).\\
G. Altarelli, R. Barbieri, Phys. Lett. {\bf B 253}, 161 (1991).\\
I. Maksymyk, C. Burgess and D. London, Phys. Rev. {\bf D50}, 529 (1994).

\bibitem{N_f}
M. Golden and L. Randall, Nucl. Phys. {\bf B361}, 3 (1990). \\
B. Holdom, J. Terning, Phys. Lett. {\bf B247}, 88 (1990).\\
S. Chivukula, M. Dugan and M. Golden, Phys. Lett. {\bf B292}, 435 (1992); Phys. Rev. {\bf D47}, 2930 (1993), hep-ph/9206222\\
M. Soldate and R. Sundrum, Nucl. Phys. {\bf B340}, 1 (1990).\\
T. Appelquist and C. Bernard, Phys. Rev. {\bf D22}, 200 (1980).\\
A. Dobado, D. Espriu, M. Herrero, Phys. Lett. {\bf B255}, 405 (1991).

\bibitem{experiments}G. Altarelli, R. Barbieri and F. Caravaglios, Int. J. Mod. Phys. {\bf A13}, 1031 (1998), hep-ph/9712368\\ 
D. Kennedy and P. Langacker, Phys. Rev. Lett. {\bf 65}, 2967 (1990). \\
G. Montagna, O. Nicrosini and F. Piccinini, hep-ph/9802302.\\
J. Erler and P. Langacker, hep-ph/9809352.

\bibitem{pdg}
D.~E.~Groom {\it et al.} (Particle Data Group),
Eur.\ Phys.\ J.\  {\bf C15}, 1 (2000). $\backslash \backslash$ pdg.lbl.gov$\backslash$



\bibitem{gasser}J. Gasser and H. Leutwyler, Ann. Phys. (N. Y.) {\bf 158}, 142 (1984); Nucl. Phys. {\bf B250}, 495 (1985).\\
J. Bijnens, G. Colangelo and J. Gasser, Nucl. Phys. {\bf B427}, 425 (1994). hep-ph/9403390.\\
J. Donoghue, E. Golowich and B. Holstein, {\em Dynamics of the Standard Model}, Cambridge U. P. 1992.

\bibitem{cqm}
 A. Manohar and H. Georgi, Nucl. Phys. {\bf B234}, 189 (1984). \newline
J. Bijnens, S. Dawson and G. Valencia, Phys. Rev. {\bf D44}, 3555 (1991).\\
D. Espriu, E. de Rafael and J. Taron, Nucl. Phys. {\bf B345}, 22 (1990); {\bf B355}, 278(E) (1991). \\
R. Ball, Chiral Gauge Theory, Phys. Rep. {\bf 182}, 1 (1989).

\bibitem{vmd}
G. Ecker, J. Gasser, A. Pich and E. de Rafael, Nucl. Phys. {\bf B321}, 311 (1989). \newline
J. Donoghue, C. Ramirez and G. Valencia, Phys. Rev. {\bf D39}, 1947 (1989).\\
G. Ecker, J. Gasser, H. Leutwyler, A. Pich and E. de Rafael, Phys. Lett. {\bf B223}, 425 (1989).  

\bibitem{tc}
S. Weinberg, Phys. Rev. {\bf D13}, 974 (1976); {\bf D19}, 1277 (1979).\\
L. Susskind, Phys. Rev. {\bf D20}, 2619 (1979).\\
E. Farhi and L. Susskind, Phys. Rep. {\bf 74}, 277 (1981).\\
R. Kaul, Rev. Mod. Phys. {\bf 55}, 449 (1983).\\
E. Farhi and R. Jackiw, {\em Dynamical Gauge Symmetry Breaking}, World Scientific 1982.\\
E. Eichten, I. Hinchleffe, K. Lane and C. Quigg, Rev. Mod. Phys. {\bf 56}, 579 (1984); {\bf 58}, 1065 (1985); Phys. Rev. {\bf D34}, 1547 (1986).

\bibitem{etc}
S. Dimopoulos and L. Susskind, Nucl. Phys. {\bf B155}, 237 (1979).\\
E. Eichten and K. Lane, Phys. lett. {\bf 90B}, 125 (1980).\\
E. Farhi and L. Susskind, Phys. Rev. {\bf D20}, 3404 (1979).\\
S. Raby, S. Dimopoulos and L. Susskind, Nucl. Phys. {\bf B169}, 373 (1980). \\
S. Dimopoulos, H. Georgi and S. Raby, Phys. Lett. {127B}, 101 (1983).\\
S. Chao and K. Lane, Phys. Lett. {\bf 159B}, 135 (1985).\\
M. Peskin, Nucl. Phys. {\bf B175}, 197 (1980).\\
J. Preskill, Nucl. Phys. {\bf B177}, 21 (1981).\\
R. Renken and M. Peskin, Nucl. Phys. {\bf B211}, 93 (1983).\\
S. Dimopoulos, Nucl. Phys. {\bf B168}, 69 (1980).

\bibitem{wtc}B. Holdom, Phys. Rev. {\bf D24}, 1441 (1981); Phys. Lett. {\bf 150B}, 301 (1985).\\
M. Bando et al., Phys. Rev. Lett. {\bf 59}, 389 (1987).\\
V. Miransky, Nouv. Cim. {\bf 90B}, 149 (1985).\\
V. Miransky and K. Yamawaki, Mod. Phys. Lett. {\bf A4}, 129 (1989).\\
R. Chivukula, and K. Lane, Nucl. Phys. {\bf B343}, 554 (1990).\\
T. Appelquist, A. Ratnaweera, J. Terning and L. Wijewardhana, Phys. Rev. {\bf D58}, 105017 (1998).\\
T. Appelquist, J. Terning and L. Wijewardhana, Phys. Rev. Lett. {\bf 77}, 1214 (1996); Phys. Rev. {\bf D44}, 871 (1991); Phys. Rev. Lett. {\bf 79}, 2767 (1997).\\
and ref. \cite{dse}

\bibitem{scaling}
R. Cahn, M. Suzuki, Phys. Rev. {\bf D44}, 3641 (1991). \\
T. Truong, Phys. Lett. {\bf B273}, 292 (1991).\\
R. Johnson, B. Young and D. McKay, Phys. Rev. {\bf D42}, 3855 (1990); {\bf D44}, E. 302 (1991).\\
S. Dimopoulos, S. Raby and G. Kane, Nucl. Phys. {\bf B182}, 77 (1981). \\
E. Eichten in ref. \cite{tc}

\bibitem{acd}
R. Sundrum and S. Hsu, Nucl. Phys. {\bf B391}, 127 (1993), p. 139, hep-ph/9206225. \\
S. Ignjatovi\'c, T. Takeuchi and L. Wijewardhana, Phys. Lett. {\bf B401}, 287 (1997), hep-ph/9702440.\\
S. Ignjatovi\'c, L. Wijewardhana and T. Takeuchi, Phys. Rev. {\bf D61}, 056006 (2000). 

\bibitem{dse}
T. Appelquist and G. Triantaphyllou, Phys. Lett. {\bf B278}, 345 (1992). \\
T. Appelquist and L. Wijewardhana, Phys. Rev. {\bf D36}, 568 (1987); {\bf D35}, 774 (1987). \\
T. Appelquist, D. Karabali and C. Wijewardhana, Phys. Rev. Lett. {\bf 57}, 957 (1986).  \\
T. Appelquist, D. Carrier, L. Wijewardhana and W. Zheng, Phys. Rev. Lett. {\bf 60}, 1114 (1988). \\
T. Appelquist, T. Takeuchi, M. Einhorn and L. Wijewardhana, Phys. Lett. {\bf B220}, 223 (1989).\\
S. King and B. Ross, Phys. Lett. {\bf B228}, 363 (1989).\\
T. Akiba and T. Yanagida, Phys. Lett. {\bf 169B}, 432 (1986). 

\bibitem{bse}
M. Harada and Y. Yoshida, Phys. Rev. {\bf D50}, 6902 (1994); {\bf D53}, 1482 (1996).  


\bibitem{gnc} 
B. Holdom and P. Lewis, Phys. Rev. {\bf D50}, 3491 (1994), hep-ph/9402319.  \\
P. Lewis, Ph. D. Thesis, U. of Toronto 1995.\\
B. Holdom, Phys. Rev. {\bf D45}, 2534 (1992). \\
J. Terning, Ph. D. Thesis, U. of Toronto 1990; Phys. Rev. {\bf D44}, 887 (1991).\\ 
B. Holdom, J. Terning and K. Verbeek, Phys. Lett. {\bf B232}, 351 (1989); {\bf B245},  612 (1990); {\bf B273}, 548 (E) (1991).\\
B. Holdom and J. Terning, Phys. Lett. {\bf B247}, 88 (1990). 

\bibitem{ENJL}
J. Bijnens, C. Bruno and E. de Rafael, Nucl. Phys. {\bf B390}, 501 (1993), hep-ph/9206236; hep-ph/9502393.\\
J. Bijnens and E. de Rafael, Z. Phys. {\bf C62}, 437 (1994),hep-ph/9306323.\\
J. Bijnens and J. Prades, Nucl. Phys. (Proc. Suppl.) {\bf 39B}, 245 (1995); hep-ph/9409231.\\
T. Hatsuda and T. Kunihiro, Phys. Rep. {\bf 247}, 221 (1994).\\
G. Nambu and G. Jona Lasinio, Phys. Rev. {\bf 122}, 345 and {\bf 124}, 246 (1961).\\
D. Ebert and H. Reinhardt, Nucl. Phys. {\bf B271}, 188 (1986).

\bibitem{negative-S}
B. Holdom, Phys. Lett. {\bf B259}, 329 (1991). \\
E. Gates and J. Terning, Phys. Rev. Lett., {\bf 67}, 1840 (1991). \\
M. Luty and R. Sundrum, Phys. Rev. Lett. {\bf 70}, 529 (1993), hep-ph/9209255. \\
B. Dobrescu and J. Terning, Phys. Lett. {\bf B416}, 129 (1998), hep-ph/9709297.\\ 
M. Dugan and L. Randall, Phys.  Lett.  {\bf B264}, 154 (1991).\\ 
M. Knecht and E. de Rafael, Phys. Lett. {\bf B424}, 335 (1998), hep-ph/9712457.\\
Z. Xiao, L. Wan, G. Lu, X. Wang and Z. Yuan, Commun. Theor. Phys. {\bf 24}, 91 (1995).\\ 
P. Bamert, C. Burgess, Z. Phys. {\bf C66}, 495 (1995), hep-ph/9407203\\ 
T. Rizzo, Phys. Rev. {\bf D50}, 2256 (1994), hep-ph/9403241.

\bibitem{critics}
K. Lane, hep-ph/9409304; hep-ph/9501249.\\
U. Mahanta, Phys. Rev. Lett. {\bf 62}, 2349 (1989). \\
K. Yamawaki, M. Bando and K. Matumoto, Phys. Rev. Lett. {\bf 56}, 1335 (1986)\\
M. Bando, K. Yamawaki and K. Matumoto, Phys. Lett. {\bf B178}, 308 (1980) \\
T. Appelquist and F. Sannino, Phys. Rev. {\bf D59}, 067702 (1999), hep-ph/9806409.\\
T. Appelquist, P. Rodrigues da Silva and F. Sannino, Phys. Rev. {\bf D60}, 116007 (1999), hep-ph/9906555.\\
A. Cohen and H. Georgi, Nucl. Phys. {\bf B314}, 7 (1989).\\
C. Bernard, A. Ducan, J. LoSecco and S. Weinberg, Phys. Rev. {\bf D12}, 792 (1975).\\
and refs. \cite{dse}

\bibitem{isospin}
P. Sikivie, L. Susskind, M. Voloshin and  V. Zakharov, Nucl. Phys. {\bf B173}, 189 (1980). \\ 
R. Chivukula, Phys. Rev. Lett. {\bf 61}, 2657 (1988).\\
T. Appelquist, T. Takeuchi, M. Einhorn and L. Wijewardhana, Phys. Lett. {\bf B232}, 211 (1989). \\
T. Appelquist and J. Terning, Phys. Lett. {\bf B315}, 139 (1993), hep-ph/9305258.\\
H. Goldberg, Phys. Rev. Lett. {\bf 58}, 633 (1987).\\
A. Falk, HUTP-89/A036 (Harvard University preprint). \\
T. Yoshikawa, Prog. Theor. Phys. Suppl. {\bf 123}, 163 (1996); Mod. Phys. Lett. {\bf A10}, (1995), hep-ph/9411280. 

\bibitem{higashijima}
K. Higashijima and A. Nishimura, Nucl. Phys. {\bf B113}, 173 (1976).\\
K. Higashijima, Phys. Rev. {\bf D29}, 1228 (1984). 

\bibitem{sumrules}
J. Donoghue and E. Golowich, Phys. Rev. {\bf D49}, 1513 (1994).\\
R. Peccei and J. Sol\'a, Nucl. Phys. {\bf B281}, 1 (1987).\\
C. Roiesnel and T. Truong, Phys. Lett. {\bf B253}, 439 (1991). \\
M. Shifman, A. Vainshtein and V. Zakarov, Nucl. Phys. {\bf B147}, 385, 488 and 519 (1979).\newline
L. Reinders, H. Rubenstein and S. Yazaki, Phys. Rep. {\bf 127}, 1 (1985).\\ 
S. Narison, {\em QCD Spectral Sum Rules}, World Scientific 1989.\\
E. Floratos, S. Narison and E. de Rafael, Nucl. Phys. {\bf B155}, 115 (1979).\\
 S. Weinberg, Phys. Rev. Lett {\bf 18}, 188 and 507 (1967).\newline
T. Das, V. Mathur and S. Okubo, {\bf 18}, 761 (1967); {\bf 19}, 859 (1967).

\bibitem{large-N}
G. 't Hooft, Nucl .Phys. {\bf B72}, 461 (1974).\\
E. Witten, Nucl. Phys. {\bf B160}, 57 (1979).\newline
S. Coleman, {\em Aspects of Symmetry}. Cambridge U. P. 1985. Chap. 8\\
S. Das, Rev. of Mod. Phys. {\bf 59}, 235 (1987).

\bibitem{lqs-bs}
C. Burden, L. Qian, C. Roberts, P. Tandy and  M. Thomson, Phys. Rev. {\bf C55}, 2649 (1997), nucl-th/9605027.\\
C. Roberts and A. Williams, Prog. Part. Nucl. Phys. {\bf 33}, 477 (1994).\\
P. Maris and C. Roberts, Phys. Rev. {\bf C56}, 3369 (1997).\\
C. Roberts, R. Cahill, M. Sevior and N. Iannella, Phys. Rev. {\bf D49}, 125 (1994).\\
C. Savkli and F. Tabakin, Nucl. Phys. {\bf A628}, 645 (1988), hep-ph/9702251.\\
C. Burden and D. Lie, Phys. Rev. {\bf D55}, 367 (1997), hep-ph/9605328. \\
P. Jain and H. Munczek, Phys. Rev. {\bf D48}, 5403 (1993); {\bf D44}, 1873 (1991).\\
D. Kekez and D. Klabucar, Phys. Lett. {\bf B457}, 359 (1999), hep-ph/9812495).\\
R. Cahill and S. Gunner, Phys. Lett. {\bf B359}, 281 (1995),  hep-ph/9506445.\\
C. Munz, J. Resag, B. Metsch and H. Petry, Nucl. Phys. {\bf A578}, 418 (1994), nucl-th/9307027.\\ 
K. Aoki, M. Bando, T. Kugo and M. Mitchard, Prog. Theor. Phys. {\bf 85}, 355 (1991) \\
F. Gross and J. Milana, Phys. Rev. {\bf D50}, 3332 (194).\\
A. Sommerer, A. El-Hady, J. Spence and J. Vary, Phys. Lett. {\bf B348}, 277 (1995), nucl-th/9412029. 

\bibitem{nrqm}
S. Godfrey and N. Isgur, Phys. Rev. {\bf D32}, 189 (1985).\\
A. Le Yaouanc, Ll. Olivier, O. Pen\'e and J. Raynal, {\em Hadron
Transitions in the Quark Model}, Gordon Breach 1988.\\
R. Van Royen and V. Weisskopt, Il Nouv. Cim. {\bf L A}, 617 (1967).

\bibitem{bs-se} 
M. Strassler and M. Peskin, Phys. Rev. {\bf D43}, 1500 (1991).\\ 
S. Caswell and G. Lepage, Phys. Rev. {\bf A18}, 810 (1978).  \\
J. Bijtebier and J. Broekaert, J. Phys. {\bf G22}, 559 (1996); Erratum-ibid. {\bf G22}, 1237 (1996). 

\bibitem{cornell} 
E. Eichten, K. Gottfried, T. Kinoshita, K. Lane and T. Yan, Phys. Rev. {\bf D17}, 3090 (1978) {\bf D21}, 313(E) (1980); {\bf D21}, 203 (1980).\\
E. Eichten, et al., Phys. Rev. Lett. {\bf 34}, 369 (1975).

\bibitem{richardson}  
J. Richardson, Phys. Lett. {\bf B82}, 272 (1979).\\
R. Levine and Y. Tomozawa, Phys. Rev. {\bf D19}, 1572 (1979). \\
W. Celmaster and F. Henyey, Phys. Rev. {\bf D18}, 1688 (1978).\\
W. Buchm\"uller and S. Tye, Phys. Rev. {\bf D24}, 132 and 3003 (1981). \\
J. Pantaleone, S. Tye and Y. Ng, Phys. Rev. {\bf D33}, 777 (1986); Phys. Rev. Lett. {\bf 55}, 916 (1985).\\
P. Moxhay and J. Rosner, Phys. Rev. {\bf D28}, 1132 (1983).\\
A. Pineda and F. Yndur\'ain, hep-ph/9711287.

\bibitem{twoloops}
S. Narison, Phys. Rep. {\bf 84}, 263 (1982) (pag. 339); Z. Phys. {\bf C2}, 11 (1979). \\
K. Schilcher, M. Tran and N. Nasrallah, Nucl. Phys. {\bf B181}, 91 (1981); Err., ibid. {\bf B187}, 594 (1981). \\
T. Chang, K. Gaemers and W. van Neerven, Nucl. Phys. {\bf B202}, 407 (1982).\\
L. Reinders, S. Yazaki and  H. Rubinstein, Phys. Rep. {\bf 127}, 1 (1985) (pag. 22).\\
B. Kniehl, Nucl. Phys. {\bf B347}, 86 (1990).\\
P. Broadhurst, Phys. Lett. {\bf B101}, 423 (1981); Z. Phys. {\bf C47}, 115 (1990).\\
F. Berends and J. Tausk, Nucl. Phys. {\bf B421}, 456 (1994).

\bibitem{ksrf}
K. Kawarabayashi and M. Suzuki ,Phys. Rev. Lett. {\bf 16}, 255 (1966).\\
Riazuddin and Fayyazuddin, Phys. Rev. {\bf 147}, 1071 (1966).

\bibitem{decoupling}T. Appelquist and J. Carazzone, Phys. Rev. {\bf D11}, 2856 (1975).\\
K. Zymanzik, Comm. Math. Phys. {\bf 34}, 7 (1973).\\
J. Collins, {\em Renormalization}, Cambridge 1984.

\bibitem{relativistic}
W. Lucha, H. Rupprecht and F. Sch\"oberl, Phys. Rev. {\bf D46}, 1088 (1992).\\
C. Semay and B. Silvestre-Brac, Phys. Rev. {\bf D46}, 5177 (1992).

\bibitem{csb}
D. Atkinson and P. Johnson, Phys. Rev. {\bf D37}, 2296 (1988). \\
A. Barducci, R. Casalbuoni, S. De Curtis, D. Dominici and R. Gatto, Phys. Rev. {\bf D38}, 238 (1988).  \\
K. Higashijima in ref \cite{higashijima}\\
P. Castorina and S. Pi, Phys. Rev. {\bf D31}, 411 (1985).\\
H. Hamber and C. Parisi, Phys. Rev. Lett. {\bf 47}, 1792 (1981).\\
J. Kogut, J. Polongy, H. Wyld and D. Sinclair, Nucl. Phys. {\bf B225} [FS9], 326 (1983); Phys. Rev. Lett. {\bf 54}, 1475 (1985).\\
Y. Nambu, Phys. Rev. Lett. {\bf 4}, 380 (1960); Phys. Rev. {\bf 117}, 648 (1960).\\
V. Miransky, {\em Dynamical Symmetry Breaking in Quantum Field Theories}, World Scientific 1993.

\bibitem{latt-pot}
S. Aoki, et al. (PC-PACS Coll.), hep-lat/9809185.\\
E. Laermann, in {\em QCD 20 years later}, eds. P. Zerwas and H. Kastrup, World Scientific 1993, pag. 455.

\bibitem{regge}T. Regge, Nuov. Cim. {\bf 14}, 951 (1959); Nuov. Cim. {\bf 
18}, 947 (1960).\\
S. Frautschi, {\em Regge Poles and S-Matrix Theory}, Benjamin N. Y. 1963.\\
E. Squires, {\em Complex Angular Momenta and Particle Physics}, Benjamin, N.
Y. 1963.\\
W. Buchm\"uller, G. Grunberg and S. Tye in ref. \cite{richardson}.

\end{references}
\end{document}